# To beer or not to beer: does tapping beer cans prevent beer loss? A randomised controlled trial.

Elizaveta Sopina[1,2*], Irina E. Antonescu[2,3,5], Thomas Hansen[2,4], Torben Højland[4], Morten M. Jensen[2,4,5], Simon V. Pedersen[2,4], Wade Thompson[2, 6], Philipp Weber[2,6], Jamie O'Halloran[1,2], Melissa G. Beach[2,8], Ryan Pulleyblank[1,2], Elliot J. Brown[5,9,10]

[1] DaCHE, Department of Public Health, University of Southern Denmark, Denmark

[2] The PhD Association at the University of Southern Denmark (PAUSD), University of Southern Denmark

[3] Department of Physics, Chemistry and Pharmacy, University of Southern Denmark, Denmark

[4] Department of Chemical Engineering, Biotechnology & Environmental Technology, University of Southern Denmark, Denmark

[5] The PhD Association Network of Denmark (PAND), Denmark

[6] Research Unit of General Practice, Department of Public Health, University of Southern Denmark, Denmark

[7] Department of Mathematics and Computer Science, University of Southern Denmark, Denmark

[8] CI2M, Centre for Integrative Innovation Management, Department of Marketing and Management, University of Southern Denmark, Denmark

[9] The PhD Association of DTU, the Technical University of Denmark, Denmark

[10] Section for Ecosystem Based Marine Management, National Institute of Aquatic Resources, the Technical University of Denmark, Denmark

* Corresponding Author

Correspondence to:

Elizaveta Sopina
J.B. Winsløws Vej 9, Odense, Denmark 5000
Phone: +45 65 50 83 84; Email: lsopina@health.sdu.dk

Authors' contributions

Elizaveta Sopina (ES) and Elliot J. Brown (EJB) devised the idea for this study, which was further developed by ES, EJB, Irina E. Antonescu (IEA), Morten M. Jensen (MMJ), Simon V. Pedersen (SVP), and Thomas Hansen (TBH). All authors contributed to the design of the experiment. IEA, SVP and TBH secured the research materials. ES, IEA, TBH, Torben Højland (TRH), Wade Thompson (WT), Philipp Weber (PW), Melissa G. Beach (MGB) and EJB conducted the experiment and collected the data. ES, Jamie O'Halloran (JOH) and EJB conducted the initial analysis of the results. All authors helped interpret the results. All authors worked on revisions of the article and approved the final version.

## Ethics committee approval
No approval was required, as no human or animal participant data was collected for this study.

## Data sharing
Data available

## Funding
No funding was received for this study. The materials for the experiment (beer cans) were provided by Carlsberg Breweries A/S, who had no vested interest in the outcome of the study and were not involved in any part of the study conception, design, analysis or manuscript writing.

# Abstract

**Objective:** Preventing or minimising beer loss when opening a can of beer is socially and economically desirable. One theoretically grounded approach is tapping the can prior to opening, although this has never been rigorously evaluated. We aimed to evaluate the effect of tapping a can of beer on beer loss.

**Methods:** Single centre parallel group randomised controlled trial. 1031 cans of beer of 330 mL were randomised into one of four groups before the experiment: unshaken/untapped (n=256), unshaken/tapped (n=251), shaken/untapped (n=249), or shaken/tapped (n=244).The intervention was tapping the can of beer three times on its side with a single finger. We compared tapping versus non-tapping for cans that had been shaken for 2 minutes or were unshaken. Three teams weighed, tapped or did not tap, opened cans, absorbed any beer loss using paper towels, then re-weighed cans. The teams recorded the mass of each can before and after opening with an accuracy of ±0.01 grams.

**Main outcome measure:** The main outcome measure was beer loss (in grams). This was calculated as the difference in the mass of the beer after the can was opened compared to before the can was opened.

**Results:** For shaken cans, there was no statistically significant difference in the mass of beer lost when tapping compared to not tapping (mean difference of-0.159g beer lost with tapping, 95% CI-0.36 to 0.04). For unshaken cans, there was also no statistically significant difference between tapping and not tapping.

**Conclusion:** These findings suggest that tapping shaken beer cans does not prevent beer loss when the container is opened. Thus, the practice of tapping a beer prior to opening is unsupported. The only apparent remedy to avoid liquid loss is to wait for bubbles to settle before opening the can.

## Introduction

Beer has been brewed by humans since prehistoric times (Meussdoerffer, 2009). In the modern era, beer is the most popular alcoholic beverage (Colen & Swinnen, 2010) and holds an important and valuable place in the social aspect of human interaction. Global beer consumption has been increasing in the past decades, reaching more than 1700 million hectolitres in 2013 (Madsen & Wu, 2016).

Beer is commonly sold and consumed from aluminium cans (Deshwal & Panjagari, 2019). Being a lightly carbonated beverage, beer is subject to 'fizzing' out of the containing vessel, especially when the vessel is shaken or disturbed (e.g. when carried in a shopping bag or on a bicycle). This is inefficient, as fizzing reduces the amount of beer available for consumption and results in waste. Beer spray can also stain clothes or surrounding objects, and therefore is also an unpleasant and socially undesirable side-effect. Therefore, preventing, or, at least, minimising beer fizzing is both socially and economically desirable.

In the manufacturing of beer, carbon dioxide ($CO_2$) is dissolved in an aqueous solution under relatively high pressure (Vega-Martínez, Enríquez, & Rodríguez-Rodríguez, 2017). If the vessel has been shaken prior to opening, $CO_2$ bubbles form in the solution, especially at points of contact with the can surface. When the vessel is depressurised by opening, the $CO_2$ becomes supersaturated and gets partly released into solution and subsequently joins the preformed bubbles, which results in rapid bubble growth (Hamlett, 2016). As the bubbles try to swiftly evacuate the vessel due to large density difference between the gas and the liquid, these $CO_2$ bubbles carry solution with them to the surface, causing the beer to overflow from the containing vessel (Arafat, 2017; Hamlett, 2016).

One proposed approach for preventing/minimising beer loss is tapping or flicking the can prior to opening. This has some theoretical grounding (Hamlett, 2016). The hypothesis is based on the assumption that tapping induces vibrations dislodging the wall-adhered bubbles. These then rise to the top of the can and possibly out of the liquid into the headspace of the can (Hamlett, 2016). However, to the best of our knowledge, the effect of tapping cans to prevent beer loss has never been rigorously evaluated.

Anecdotes and opinion pieces exist on whether tapping carbonated beverages reduces liquid loss and seem to suggest tapping is not effective (Arafat, 2017; Gammon, 2011; Lane, 2007; Mikkelson, 2007). However, these pieces are not based on well-designed experiments, focus mainly on soda, and include input from the big soda industry who may have a conflict of interest in this matter. Since expert opinion is considered the lowest level of the evidence (Balshem et al., 2011), or sometimes does not even feature in the evidence pyramid (Murad, Asi, Alsawas, & Alahdab, 2016), we should not be relying on these pieces as definitive evidence in this area.

Given the strong Danish tradition in beer brewing (Glamann, 1962) and consumption, and an emerging trend in exploring alcohol-related urban myths (Hansen, Færch, & Kristensen, 2010), we set out to settle this matter with high-quality evidence. The aim of this paper was to investigate whether tapping a beer can prior to opening prevents beer loss.

## Methods

We conducted a randomised controlled trial with 1031 330 mL cans of 'Pilsner'-style beer. Using Microsoft Excel's random number generator, cans were randomised into one of four groups: (1) unshaken/untapped, (2) unshaken/tapped, (3) shaken/untapped, (4) shaken/tapped. The sample size was selected as the result of power calculations based on a small pilot study, where an effect size of 3 grams was observed.

The day prior to the experiment, cans were individually labelled using large numbered stickers. According to the random assignment of the can number to one of the four treatment groups, two sets of colour-coded stickers were applied to represent shaken/non-shaken (pink or purple) and tapped/non-tapped (yellow or orange). Shaken/non-shaken labels were placed on the bottom of the can and tapped/non-tapped stickers were placed on the side of the can. The cans were subsequently refrigerated at 4°C overnight.

The experiment took place on Friday, 18th of May 2018 between 14:20 and 19:50 at the Technical Campus of the University of Southern Denmark, Odense, Denmark. During the experiment, cans were taken out of the fridge in batches of 20 to 30 cans and kept in polystyrene containers to keep them cool, until being placed on shakers (if randomised to be shaken). Cans randomised to the shaking group were then clamped onto the Heidolph Instruments Unimax 2010 shaker and shaken for 2 minutes at 440 rpm in batches of three to four cans. Pilot testing revealed that this shaking method successfully mimicked carrying beer on a bicycle for 10 minutes – a common way of transporting beer in Denmark. Once ready for opening, all cans were passed to the weighing and opening teams (Figure 1).

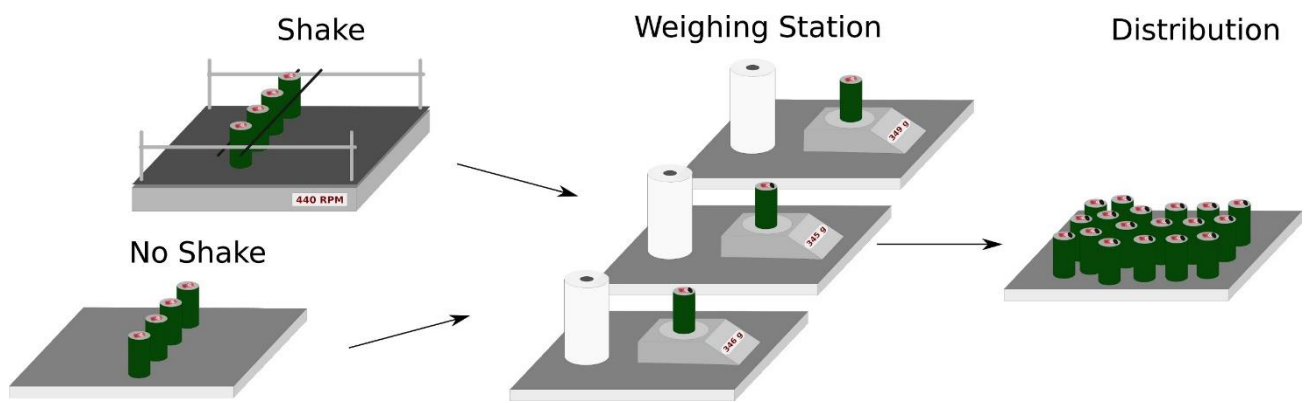

**Figure 1: Experimental design**. Cans randomised to the shaking group are clamped onto the Heidolph Instruments Unimax 2010 shaker and shaken with 440 rpm for two minutes in batches of three and four. Once ready for opening, both shaken and still cans are passed to the weighing and opening station. Here, the cans' initial weights are registered and they are tapped if required. Then the cans are opened and the spillover beer is absorbed by paper towels. After opening, the weight is registered again and the cans are passed to the distribution point.

Three teams of two wiped, weighed, tapped (if required) and opened the cans. These teams were blind to which cans were shaken (as the identifying sticker was located on the bottom of the can). One person on each team recorded the number and mass of the can before opening, the time the can was opened, and the mass after opening. The team members responsible for weighing and recording were blind to both the shaken and tapped status of the can. The mass was recorded to the nearest 0.01 grams. The other team member wiped and tapped the can (if randomised for tapping), opened it, and absorbed any beer loss using paper towels. Tapping was limited to three single-finger flicks on the side of the can. We refer to this practice as 'tapping' in order to be consistent with the established terminology (Arafat, 2017; Hamlett, 2016; Lane, 2007; Mikkelson, 2007). Each can was tapped in the same place - the recycling label printed on the side of the can. To maintain consistency, only one of the team members tapped the can throughout the experiment.

**Statistical analysis**

The primary outcome was the difference in the mass of beer after the can was opened compared to before (amount of beer lost). We compared the amount of beer lost for untapped versus tapped beers (for both shaken and unshaken beers) using generalised linear models (GLMs). Shaken and not shaken cans were analysed in two different models. The response variable (beer loss in grams) was modelled as a function of tapping/not tapping, team, batch number (for both models) and time from shaking in seconds (for shaken cans only). Quantile plots and Akaike information criterion (AIC) and Bayesian information criterion (BIC) (Bozdogan, 1987) were used to assess the performance of Gaussian, Poisson and Gamma distribution families and their associated link functions in GLMs. The non-shaken group was modelled using a Gamma distribution and a log link. For ease of interpretation, coefficients and confidence intervals have been presented in an exponentiated form where the log link function was used. A normal distribution with an identity link function was selected for the shaken group. All data analysis was conducted in Stata version 15 (StataCorp, 2015).

# Results

We processed the cans of beer in 135 batches with the mean batch size of 8, half of which were shaken and half unshaken, according to the predetermined randomisation. After excluding individual observations due to data collection errors (damaged cans, errors in shaking/tapping and duplicate records) we were left with a sample size of 1000 cans (97% of the original sample), of which 507 were not shaken controls, and 493 were in the shaken group. These exclusions resulted in 50% of the not shaken cans being allocated to the intervention (tapping), and a slightly reduced 49% of the shaken cans being administered tapping.

The initial masses of the cans between the shaken and not shaken groups were not significantly different (t = 1.73, p > 0.05), with only 0.55g difference at baseline (Table 1). As expected, the cans which were subject to shaking lost more mass than cans in the not shaken group. The effect of shaking was a mean loss of 3.45 grams loss of beer upon opening, whereas the not shaken cans only lost 0.51 grams, equating to a 6 times greater loss of beer for cans subject to shaking (Figure 2). The mean time from shaking to intervention and opening was 94 seconds (SD = 57.90, n = 1000), with a maximum of 382 and a minimum of 15 seconds.

| Table 1: baseline characteristics | | | | | |
|---|---|---|---|---|---|
| | **Not Shaken** | | **Shaken** | | **Total** |
| **Number of cans (%)** | 507 (51) | | 493 (49) | | 1000 |
| **Tapped, number (%)** | 251 (50) | | 244 (49) | | 495 (50) |
| **Can mass before opening, grams, mean (±SD)** | 349.31 (5.04) | | 348.76 (5.01) | | 349.05 (5.03) |
| **Can mass after opening, grams, mean (±SD)** | 348.80 (4.61) | | 345.32 (4.68) | | 347.08 (4.96) |
| **Mass difference, grams, mean (±SD)** | -0.51 (1.00) | | -3.45 (1.13) | | -1.96 (1.82) |
| **Time from shaking to opening, seconds, mean (±SD)** | | | 94.63 (57.90) | | |
| **Number of cans opened by each team** | **Tapped** | **Not tapped** | **Tapped** | **Not tapped** | |
| **Team 1, number (%)** | 73 (26) | 73 (26) | 71 (25) | 64 (23) | 281 |
| **Team 2, number (%)** | 75 (23) | 75 (23) | 82 (25) | 97 (29) | 329 |
| **Team 3, number (%)** | 103 (26) | 108 (28) | 91 (23) | 88 (23) | 390 |

Of the three teams tasked with administering the intervention and data collection, each team opened a similar ratio of not shaken to shaken cans. Team three processed the highest number of cans in total (390 cans). Overall there were no statistically significant differences in the characteristics between the non-shaken and shaken cans.

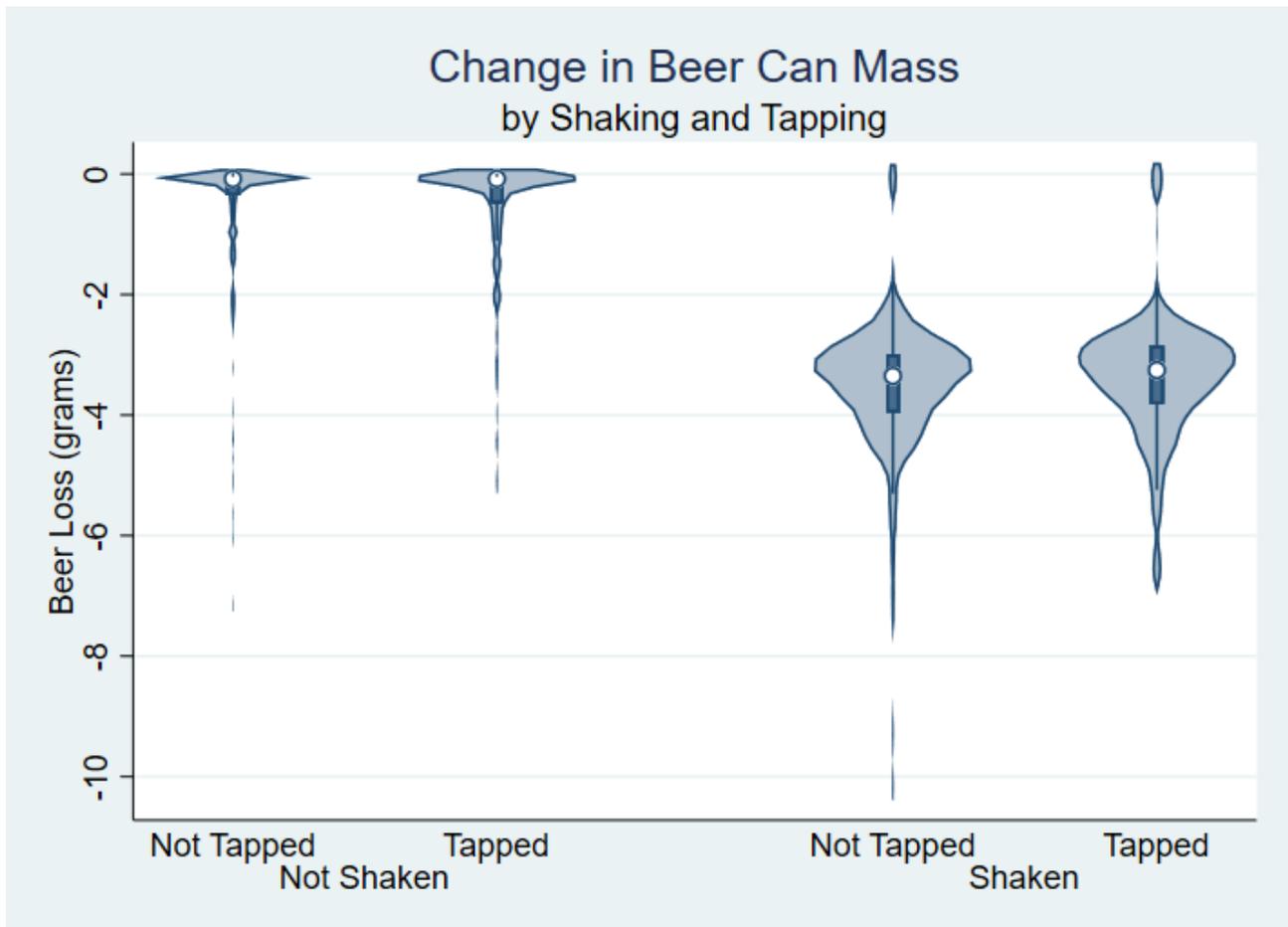

**Figure 2: Change in beer can mass by shaking and tapping.** The density traces illustrate the mode at their widest point, while the internal solid objects represent the quartiles (box), median (○), and 1.5 times the interquartile range (whiskers) of a traditional box plot.

GLM analyses revealed no statistically significant difference in beer loss between tapped and non-tapped cans, for shaken (mean difference 0.159g less beer lost for tapped versus untapped, 95% CI -0.36 to 0.04) and not shaken cans (mean difference 0.085g more beer lost for tapped versus untapped, 95% CI, -0.22 to 0.50 g) (Table 2).

Our results revealed a statistically significant team effect on the amount of beer lost. Team 3 had a mean reduction in the lost beer of 0.51g and 0.58g for non-shaken and shaken groups, respectively (Table 2), compared to the next best group (*p*<0.01).The batch number had a statistically significant, but small effect of beer lost (0.005 grams, *p*<0.01) in the non-shaken group. Finally, we also found that the time from shaking to the intervention had no statistically significant effect on the amount of beer lost.

| VARIABLES | Not Shaken | Shaken |
|---|---|---|
| **Table 2: GLM findings: the effect of tapping and other factors on the loss of beer from cans (g)** | | |
| **Tapping** | 0.085 (-0.216 - 0.503) | -0.159 (-0.363 - 0.0449) |
| **Team (Ref: Team 1)** | | |
| **Team 2** | 0.080 (-0.298 - 0.661) | -0.154 (-0.415 - 0.106) |
| **Team 3** | -0.512*** (-0.672 - -0.274) | -0.579*** (-0.847 - -0.312) |
| **Time from shaking (seconds)** | | -0.002 (-0.003 - 0.001) |
| **Batch** | -0.005*** (-0.009 - 0.000) | -0.001 (-0.005 - 0.001) |
| **Constant** | -0.204 (-0.574 - 0.485) | 4.152*** (3.677 - 4.626) |
| **Observations** | 496 | 454 |

Confidence intervals in parentheses
 *** p<0.01

## Discussion

**Reasons for intervention (in)effectiveness**

This study answers the question of whether tapping beer cans preserves the precious liquid when the container is opened. The results of this randomised control trial on 1000 beers do not support this hypothesis. We propose four possible reasons why the hypothesis was disproved.

First, the energy of the flicking provides only negligible energy dissipation as a driving force to relocate an adequate proportion of bubbles to the headspace. This may be due to the energy of the flick being absorbed by the aluminium and the bulk liquid.

Secondly, the radius of the bubbles is under a specific threshold that might allow the bubbles to ascend fast enough to the top of the can. We also had a relatively low waiting time from tapping to opening, which may have resulted in a reduced time interval for any dislodged bubbles to rise.

Thirdly, the premise of the hypothesis is that beer spillage is primarily driven by the turbulent surfacing of wall-adhered bubbles. However, if most bubbles are located in the bulk liquid, the surfacing of the wall-adhered bubbles by flicking would be insignificant compared to the rapid surfacing of the bubbles in the bulk liquid. This is consistent with the tendency shown in Figure 2, where the tapped cans show no statistically significant reduction of spillage compared to non-tapped cans. This is not surprising given the experimental design for processing cans from shaking to intervention, opening and data recording was smooth and relatively fast. The average time from shaking to the intervention was 58 seconds, with a minimum of 15 and a maximum of 382. It is probable that the variation in time from shaking to opening was not significant for any of the 'fizziness' to subside. If the result of tapping was evident within this time-scale, it is likely that our results would demonstrate this effect. Hence, if tapping the can should give a statistically significant reduction in beer loss, it would only be meaningful if it occurs at time-scales shorter than the one explored in the present study.

Finally, another aspect that could explain the absence of beer loss reduction is the composition of beer. Beer contains barley proteins, *i.e.* lipid transfer protein 1 (Siebert, 2014) and protein Z (Niu et al., 2018) which are known to contribute to the "creaminess" of beer foam and have a stabilizing effect on the internal microbubbles formed due to shaking. These proteins may prevent the microbubbles from rising to the top of the can after tapping. The presence of these components explains why the foam head is more stable for beer than for carbonated drinks that lack these foam-stabilising proteins. Moreover, the foaming in carbonated beverages also depends on the surface tension and density of the respective aqueous solution. Therefore, we cannot assume an identical result to a soda experiment.

As barley proteins in native conformation stabilise the surface of bubbles in the liquid, and thus increases potential overflow, a strategy for reducing this risk could be the denaturation of proteins naturally occurring in beer. Different strategies may be employed in the denaturation of the protein component of beer *e.g.* adding chaotropic agents, ultra-sonication, thermal denaturation, radiation etc. Since most of these methods involve the opening of the cans (*e.g.* adding chaotropic agents), expensive equipment (*e.g.* ultra-sonication) or potential health risk, either during the denaturation process or the following consumption of the beverage (*e.g.* radiation), heat denaturation appears to be the most applicable method. However, the bubbles are also temperature-sensitive in the way that an increase in temperature will cause an increase in the amount of overflow (Sahu, Hazama, & Ishihara, 2006). This introduces a prolonged wait step to lower the temperature again, both to reduce the temperature effect on bubbles, and to reach a beer temperature that most people would enjoy. Thus, a possible approach to reduce the risk of overflow in beer consumption scenarios, where shaking is introduced before consumption/opening the can, is to introduce a heating step prior to the usual cooling step beer consumers apply. The effect of heat denaturation, its potential negative impact on the sensory experience of the beer consumption and the risk of applying heat to a sealed pressurized metal container are important future research topics to be answered.

### Broader social implications

There are a number of broader socially-relevant concerns associated with excess beer loss due to fizzing. The potential for tapping-related finger injuries (*e.g.* damaged tendons, ligaments, repetitive strain disorder) (Leggit & Meko, 2006) can be minimised by disseminating the strong evidence of no effect. The full extent of the associated healthcare costs and productivity loss due to tapping-related injuries is unknown but considering these crucial factors in future research would arm public health experts with an additional reason to integrate the message that beer tapping is not an effective method of reducing beer fizzing. Furthermore, enthusiastic beer consumers should be targeted as a priority group as they are at the highest risk of tapping-related injuries.

Excessive alcohol consumption is an acknowledged global health concern (World Health Organization, 2014). From a public health perspective, another important message would be that beer loss is minimised when the can is settled (not shaken). Therefore, beer drinkers who might mistakenly consider tapping cans to be a way to reduce beer loss may be convinced to slow down drinking while letting the beer settle. Promoting the message that slowing the rate of consumption minimises waste, it is conceivable to suggest both improvements in the efficiency of beer consumption and reduction in alcohol-related harm.

It is worth noting that the absolute amount of beer lost for a shaken beer may appear fairly low (around 3.5 grams). There is likely to be individual variability around how much importance is placed on this amount of lost beer. Our study suggests that one whole can of beer can be preserved by allowing approximately 100 shaken cans to settle. Post-secondary students, economists and other frugal beer enthusiasts are likely to find satisfaction in this fact.

## Future research directions

Since tapping does not affect beer loss from opening shaken cans, the only known remedy to avoid spillage is to wait for the bubbles to settle before opening the can. The amount of time required for the bubbles to settle depends on several factors, including the rigour of shaking and characteristics of the aqueous solution such as density, viscosity and surface tension. An interesting next step would be to determine the required settling time for a broad spectrum of carbonated drinks. In addition, less vigorous shaking may produce a higher percentage of wall-adsorbed bubbles susceptible to elimination by tapping. This may elucidate whether tapping under the right conditions can significantly reduce spillage. It may also be of interest to test whether tapping a different part of the can, for example, the top would produce different results.

## Strengths and limitations

A number of tests on the effectiveness of tapping cans to prevent beverage loss have been conducted, although these were often not conducted in a scientific manner or subject to peer-review, reporting their findings in the grey literature (Arafat, 2017; Hamlett, 2016; Lane, 2007; Mikkelson, 2007). This study is, to the best of our knowledge, the first to apply the 'gold standard' for experimental design (Barton, 2000), and makes use of a large sample size of 1000 cans to address this question. The large sample size and the robust experimental design of this study make the findings reported in this paper reliable.

During the experiment, Team 3 had significantly lower beer loss from both shaken and unshaken cans, compared to the other teams. The most likely explanation for this is that, unlike other teams, Team 3 was exclusively comprised of engineers, who are known to go to great lengths for getting the last drop of beer.

During the experiment, it quickly became apparent that the continuous opening of cans was causing finger and nail-bed pain for researchers who were opening cans. To combat this, the use of a tool (stainless steel butter knife) to open the cans was employed. This had some positive effects on the experiment. Firstly, it helped avoid the can-opening researchers having to change fingers during the experiments, leading to different mechanics of beer opening. And secondly, it helped prevent any excessive hand and wrist injuries for the researchers.

Furthermore, the shaking of the cans, though consistent, was rather vigorous, which may have induced a higher concentration of bubbles compared to what a can would contain under non-laboratory conditions. And while the shaking method may be consistent with the vibrations a six-pack may experience in a bicycle basket, these may be more vigorous than cans transported by other means. This, combined with the short time between shaking and opening of the cans, may have contributed to a higher ratio of free-flowing bubbles in the liquid phase, to those adsorbed on the can walls (though our pilot experiments demonstrated that the shaking system mimicked the natural shaking occurring on a bicycle ridden for ten minutes on a bumpy road). With the hypothesis that the purpose of the tapping was to release bubbles from the walls, a can with a higher ratio of free- to wall-adhered bubbles would show a lesser effect from the tapping. This also raises the question of the time from the tapping to the opening of the can. If the bubbles are not allowed to rise out of the liquid phase prior to the can being opened, then the overflow of beer would be assumed equal between the two scenarios.

Finally, a more clearly defined procedure for tapping, opening and wiping down cans could help account for the natural variation in beer opening and ensure further consistency across different teams. Wiping any condensed liquid on the outside of the can before and after each weighing, regardless of the amount of visible spillage, is a possible method for decreasing variations not associated with the shaking and tapping.

It is also important to acknowledge that the aforementioned issues may affect the balance between the consistency of the experiment and the link to real life beer consumption scenarios. During the design of the experiment, the idea of using an automated can tapper and opener was proposed but finally rejected, as the aim of the research was to determine if this effect would be observable in a realistic use case, not in an over-engineered laboratory setting.

## Conclusion

We conducted a large randomised controlled trial to examine the effect of tapping the sides of beer cans to reduce beer loss through reducing 'fizzing' and found no evidence to support the hypothesised beer-saving effect of tapping. This finding has important implications for the efficiency in beer consumption, as well as timing with which beer can be consumed to minimise loss and, potentially, alcohol-related harm. Furthermore, while the null result appears robust for this particular beer and container combination, it does not eliminate the possibility of the intervention being effective on other beverages and containers.

**Acknowledgements:**

We would like to thank and acknowledge Carlsberg Breweries A/S, for providing us with the 1000+ cans of beer for this experiment; SDU TEK for making available the necessary facilities and resources to conduct this experiment; Michael Magee and Alfonso Pizarro Ramirez for their assistance in resources, preparation for the experiment, crowd and social media control during the event, distribution of beer and communication of the project idea to event attendees; former and current members of PAUSD, for contributing to the ideas for this experiment, and general support: Nadine Heidi Brueckmann, Liubov Vasenko, Vella Somoza Sanchez, Stine Piilgaard Porner Nielsen, Rojin Kianian, Manuella Lech Cantuaria, Sergejs Boroviks, Christian Christensen, and Aske Nielsen; Hanne Kirsten Højland for help with preparing for the experiment; Katrine Astrup